\documentclass[a4paper,11pt]{article}
\usepackage{pos}

\usepackage{autobreak}
\usepackage{times}
\usepackage{epsfig,graphicx}
\usepackage[caption=false]{subfig}
\usepackage[english]{babel}
\usepackage[utf8]{inputenc} 
\usepackage{amsmath} 
\usepackage{xspace} 
\usepackage{url} 
\usepackage{slashed} 
\usepackage{framed}
\usepackage{physics}

\newcommand{\df}{{\rm d}}

\newcommand{\xbar}{\bar{x}}
\newcommand{\als}{\alpha_S}
\newcommand{\eps}{\epsilon}

\newcommand{\ns}{{\slashed{n}}}
\newcommand{\nsb}{{\slashed{\bar{n}}}}
\newcommand{\taubar}{\bar{\tau}}

\newcommand{\tkl}{t_{kl}}

\newcommand{\tl}{t_{l}}
\newcommand{\tk}{t_{k}}

\title{Automated Calculation of Beam Functions at NNLO}

\author{Guido Bell}
\author{Kevin Brune}
\author*{Goutam Das}
\author{Marcel Wald}

\affiliation{Theoretische Physik 1, 
Center for Particle Physics Siegen, Universit\"at Siegen,\\
Walter-Flex-Strasse 3, 57068 Siegen, Germany}

\emailAdd{bell@physik.uni-siegen.de}
\emailAdd{brune@physik.uni-siegen.de}
\emailAdd{goutam.das@uni-siegen.de}
\emailAdd{marcel.wald@uni-siegen.de}

\abstract{
  We present an automated framework for the calculation of beam functions that describe collinear 
  initial-state radiation at hadron colliders at next-to-next-to leading order (NNLO) in perturbation 
  theory. By exploiting the infrared behaviour of the collinear matrix elements, we factorise 
  the phase-space singularities with suitable observable-independent parametrisations. Our 
  numerical approach applies to a large class of collider observables, and as a check of its 
  validity, we compute the quark beam functions for transverse-momentum 
  resummation and N-jettiness, which are known analytically at this order, finding 
  excellent agreement. 
}

\FullConference{%
  Loops and Legs in Quantum Field Theory - LL2022,\\
  25-30 April, 2022\\
  Ettal, Germany\\[1em]
  Preprint numbers: SI-HEP-2022-18, P3H-22-085.
}

\begin{document} 
\maketitle

\section{Introduction}
\noindent
Beam functions constitute a key ingredient in factorisation theorems at hadron colliders. 
For the measurement of a global observable $\omega$ that is sensitive to soft and collinear 
QCD radiation, the differential cross section typically takes the following schematic form,
\begin{align}
  \frac{\df \sigma}{\df \omega}  = H(Q) \cdot  
            \prod_{i} B_{i/h}(\omega) \otimes 
            \prod_{j} J_{j}(\omega) \otimes 
            S(\omega) \,.
\end{align}
The hard function $H(Q)$  describes the virtual corrections to the Born process, and it depends 
only on the hard scale $Q$ of the process, while it is  independent of the specific \mbox{measurement 
$\omega$}. On the other hand, the beam functions $B_{i/h}(\omega)$, the jet functions $J_{j}(\omega)$ 
and the soft function $S(\omega)$ describe initial-state collinear, final-state collinear and 
soft emissions, respectively, and they are observable-dependent. One thus needs to compute 
these functions on a case-by-case basis for each observable, which for the beam functions has been 
achieved either 
analytically~\cite{Catani:2013tia,Gehrmann:2014yya,Luo:2019szz,Ebert:2020yqt,Luo:2020epw,Gaunt:2014xga,Gaunt:2014cfa,Behring:2019quf,Ebert:2020unb,Gaunt:2014xxa,Gaunt:2020xlc} 
or semi-analytically~\cite{Gangal:2016kuo,Abreu:2022zgo} 
in some cases relevant e.g.~for transverse-momentum or jet-veto resummation. An automated approach 
for the calculation of beam functions at next-to-next-to-leading order (NNLO) in perturbation theory 
has been initiated only recently~\cite{Bell:2021dpb,Bell:2022nrj}, and in this article we report on 
the status of these developments.

An automated framework for the calculation of soft functions at NNLO is already available through 
the public package {\tt SoftSERVE}~\cite{Bell:2018vaa,Bell:2018oqa,Bell:2020yzz}. This has been 
achieved by introducing suitable phase-space parametrisations to factorise the divergences of the 
soft matrix elements at the second order of the strong coupling. The singularity structure of the 
collinear matrix elements is, on the other hand, significantly more complicated. The beam functions 
are, moreover, defined as proton matrix elements of collinear field operators, and they need to be 
matched onto the standard parton distribution functions to extract the relevant perturbative 
information. By employing suitable phase-space parametrisations, sector-decomposition techniques 
and non-linear transformations, we recently set up a similar automated framework 
for the calculation of the beam-function matching kernels~\cite{Bell:2021dpb}. As a first application 
of our  approach, we computed the quark beam function for jet-veto resummation~\cite{Bell:2022nrj}, 
and in this work we present our results for transverse-momentum resummation and the event-shape 
variable N-jettiness. 
While these beam functions are known analytically at the considered NNLO for 
quite some time~\cite{Gehrmann:2014yya,Gaunt:2014xga}, they provide important reference observables 
for our setup. In particular, they allow us to test the numerical accuracy of our predictions 
for different classes of observables, which are known as SCET-1 and SCET-2 beam functions.

\section{Quark beam functions}
\noindent
We are concerned with quark beam functions that are defined via
\begin{align}
\label{eq:definition}
\frac12  \left[\frac{\ns}{2}\right]_{\beta \alpha}
{\cal B}_{q/h}(x,\tau,\mu) =& \sum_{X} \,
\delta\Big( (1-x) P^- - \sum_i k_i^- \Big)\,
{\cal M}(\tau;\{k_i\}) \bra{h(P)}\bar{\chi}_{\alpha} \ket{X} 
\bra{X}\chi_{\beta}  \ket{h(P)} ,
\end{align}
where $\chi = W^{\dagger}_{\bar n}\frac{\ns \nsb}{4} \psi$ is the collinear quark field, and we used 
light-cone coordinates with $k_i^- = \bar n \cdot k_i$, $k_i^+ = n \cdot k_i$ and a transverse component 
$k_i^{\perp,\mu}$ that satisfies $n \cdot k_i^{\perp}=\bar n \cdot k_i^{\perp}=0$, along with 
$n^2=\bar n^2=0$ and $n\cdot \bar n =2$. The sum over $X$ represents the phase space of the collinear 
emissions with momenta $\{k_i\}$, while the external state $\ket{h(P)}$ refers to a hadronic state 
with momentum $P^\mu=P^- n^\mu/2$. The function ${\cal M}(\tau;\{k_i\})$ furthermore specifies the observable, 
and in order to avoid distribution-valued expressions, we assume that it is given 
in Laplace space, with $\tau$ being the corresponding Laplace variable 
(see also~\cite{Bell:2018vaa,Bell:2018oqa,Bell:2020yzz}).

Unlike soft and jet functions, beam functions are intrinsically non-perturbative objects, but as long 
as the relevant scale of the collinear emissions is perturbative, i.e.~$\tau\ll1/\Lambda_{\rm QCD}$,  
they can be matched onto the usual parton distribution functions (pdf). The matching 
relation is most conveniently expressed in Mellin space,
${\cal \widehat B}_{q/h}(N,\tau,\mu) = \int_0^1 dx \;
x^{N-1} \;{\cal B}_{q/h}(x,\tau,\mu)$,
where it becomes
\begin{align}
\mathcal{\widehat B}_{q/h}(N,\tau,\mu) &=\sum_k 
\,
{\cal \widehat I}_{q\leftarrow k}(N,\tau,\mu) 
~ \widehat f_{k/h}(N,\mu)\,,
\end{align}
and the sum runs over all partonic channels. The development of an automated framework to compute the 
matching kernels ${\cal \widehat I}_{q\leftarrow k}(N,\tau,\mu) $ to NNLO accuracy is the goal of the 
current project. The matching kernels can, in fact, be extracted from partonic rather than hadronic 
beam functions, i.e.~the external states $\ket{h(P)}$ in \eqref{eq:definition} can be interpreted as 
partonic states for this purpose. If the matching is performed on-shell in dimensional regularisation, 
the partonic pdf evaluate to $\widehat f_{k/j}(N,\mu)=\delta_{kj}$ to all orders in perturbation theory, 
and the extraction of the matching kernels boils down to the calculation of the bare partonic beam functions.

Depending on the observable, dimensional regularisation may not be sufficient to resolve all 
phase-space singularities that appear in the beam-function calculation. It is well-known that one needs 
an additional prescription to regularise rapidity divergences for transverse-momentum dependent SCET-2 
observables. In our approach, we regularise these divergences with a symmetric version of the 
phase-space regulator proposed in~\cite{Becher:2011dz}. We thus introduce the following phase-space 
factor for each emission with momentum $k_i^\mu$,
 \begin{equation}
\int d^dk_i \; \left(\frac{\nu}{k_i^- + k_i^+}\right)^\alpha \;  \delta(k_i^2) \, \theta(k_i^0) \,,
\end{equation}
where $\alpha$ is the rapidity regulator and $\nu$ the corresponding rapidity scale. The choice 
of a symmetric regulator under $n \leftrightarrow\bar{n}$ exchange enables us to derive the 
anti-collinear beam function directly from the collinear one. For consistency one then also has to 
calculate the soft function in the same regularisation scheme, for which we make use of  {\tt SoftSERVE}.
 
We finally expand the bare matching kernels in the renormalised strong coupling $\als$ as
\begin{align}
\label{eq:barekernel}
&{\cal \widehat I}_{q\leftarrow k}^0(N,\tau,\nu) = \delta_{qk} 
+ \bigg(\frac{Z_\alpha\alpha_s}{4\pi}\bigg) \, 
\big( \mu^2 \taubar^2 \big)^{\eps} \;
 \bigg( \frac{\nu}{q_-} \bigg)^{\alpha} \, {\cal \widehat I}_{q\leftarrow k}^R(N,\eps,\alpha) 
\\
& \quad 
+ \bigg(\frac{Z_\alpha\alpha_s}{4\pi}\bigg)^2 \, 
\big( \mu^2 \taubar^2 \big)^{2\eps} 
 \; \Bigg\{ \bigg( \frac{\nu}{q_-} \bigg)^{\alpha} \,{\cal \widehat I}_{q\leftarrow k}^{RV}(N,\eps,\alpha)  + 
   \bigg( \frac{\nu}{q_-} \bigg)^{2\alpha} \,{\cal \widehat I}_{q\leftarrow k}^{RR}(N,\eps,\alpha) \Bigg\} + \mathcal{O}(\alpha_s^3)\,,
   \nonumber
\end{align}
where $\eps=(4-d)/2$ is the dimensional regulator,  $\bar\tau=\tau e^{\gamma_E}$, and 
$Z_\alpha = 1-\beta_0\alpha_s/(4\pi\eps)$ is the coupling renormalisation factor in the 
$\overline{\text{MS}}$-scheme. We furthermore traded the large component $P_-$ in the 
beam-function definition by the component $q_-=x P_-$ that enters the hard interaction, 
before performing the Mellin transformation. We note that the rapidity regulator $\alpha$ is 
only required for SCET-2 observables and it can thus be set to zero in the SCET-1 case.

\section{NLO calculation}
\noindent
As purely virtual corrections are scaleless and vanish in our setup, only the real-emission 
process contributes at NLO. Denoting the momentum of the emitted parton by $k^\mu$, the 
on-shell condition along with the delta function in \eqref{eq:definition} fixes two 
light-cone components to $k^+=|\vec{k}^\perp|^2/k^-$ and \mbox{$k^- = (1-x) P^-$}. 
The remaining components are then parametrised as
\begin{align}
\label{[eq:param:oneemission]}
k_T = |\vec{k}^\perp| \,, 
\qquad\qquad t_k = \frac{1-\cos \theta_k}{2} \,,
\end{align} 
where $\vec{k}^\perp$ is the transverse part of $k^{\mu}$ in light-cone coordinates, and 
$\theta_k$ is its angle with respect to a reference vector $\vec{v}^{\perp}$ in the transverse 
plane.

In terms of these variables, we write the one-emission measurement function in the form 
(see also~\cite{Bell:2018vaa,Bell:2018oqa,Bell:2020yzz}), 
\begin{align}
{\cal M}_1(\tau;k) 
&= 
\exp 
\left[ 
-\tau k_T 
\left( 
\frac{k_T}{(1-x) P^-}
\right)^n
f(t_k)
\right].
\end{align}
The observable is thus characterised by the parameter $n$, which controls the scaling of the 
observable in the soft-collinear limit~\cite{Bell:2018oqa}, and its azimuthal dependence is 
described by the function $f(t_k)$. With this ansatz one can derive a master formula for the 
calculation of the NLO matching kernels,
\begin{align}
\label{[eq:NLO]}
 {\cal \widehat I}_{q\leftarrow q}^R(N,\eps,\alpha) 
=& 
\left(\tau q_-\right)^{\frac{-2n\eps}{1+n}}
\frac{8 e^{-\gamma_E\eps}}{(1+n)\sqrt{\pi}} \,
\frac{\Gamma\left( -\frac{2\eps}{1+n}\right)}{\Gamma\left(\frac{1}{2}-\eps\right)}
\\
&\times\;\int_0^1 \!dx \;\,
x^{N + \frac{2n\eps}{1+n}+\alpha} \; \bar x^{-1 - \frac{2n\eps}{1+n}-\alpha} \;
\left[\xbar\,  \mathbb{P}_{q \to q^* g}^{(0)}(x)  \right]\;
\int_0^1 \! dt_k \;
(4 t_k \bar{t}_k)^{-\frac{1}{2}-\eps}\;
f(t_k)^{\frac{2\eps}{1+n}}\,,
\nonumber
\end{align}
where we introduced the notation $\xbar=1-x$, etc. The off-diagonal kernel
${\cal \widehat I}_{q\leftarrow g}^R(N,\eps,\alpha)$ takes a similar form with
$\mathbb{P}_{q \to q^* g}^{(0)}(x)$ replaced by $\mathbb{P}_{g \to q^* \bar q}^{(0)}(x)$. 
The latter quantities are related to the well-known (crossed) splitting functions, and they are given by
\begin{align}
\mathbb{P}_{q \to q^* g}^{(0)}(x)
&=
\frac{C_F}{x\bar x}
\left[
1+ x^2 - \eps \xbar^2
\right] \,,
\qquad
\mathbb{P}_{g \to q^* \bar q}^{(0)}(x)
=
\frac{T_F}{(1-\eps)x}
\left[
x^2+ \xbar^2 - \eps 
\right] .
\end{align}
From the master formula in \eqref{[eq:NLO]} we read off that the rapidity regulator $\alpha$ is only 
required for the diagonal channel to control the divergence in the limit $x\to 1$ as long as $n=0$, which 
corresponds to the SCET-2 case. 

\section{NNLO calculation}
\noindent
At NNLO there are two different types of contributions, \textit{viz.}\ the real-virtual (RV) and the 
real-real (RR) contribution. In the former case, the calculation follows along the lines outlined 
above, with additional explicit divergences coming from the one-loop corrections to the splitting 
functions~\mbox{\cite{Kosower:1999rx,Bern:1999ry,Sborlini:2013jba}}. In particular, the phase-space
 divergences can still be exposed with the parametrisation given in \eqref{[eq:param:oneemission]}, 
 and one can derive a similar master formula as for NLO contribution.

The RR case, on the other hand, is more complicated. It involves two emissions with momenta $k^\mu$ 
and $l^\mu$, which we parametrise 
according to
\begin{align}
\label{eq:parametrisation:two}
a = \frac{k^- l_T}{l^- k_T}, \qquad
b = \frac{k_T}{l_T}, \qquad
z = \frac{k^- + l^-}{P^-}, \qquad
q_T = \sqrt{(k^- + l^-)(k^+ + l^+)}\,,
\end{align}
where again $k_T = |\vec{k}^\perp|$ and $l_T = |\vec{l}^\perp|$. In physical terms, the variable $a$
represents a measure of the rapidity difference of the two emissions, $b$ is the ratio of their 
transverse components, and $z$ is the splitting variable of the joint system composed out of the two 
emitted particles. Notice that $q_T$ is the only dimensionful variable in this parametrisation, whereas 
there are further angular variables that we parametrise similar to the NLO case from above. Two of 
these angular variables ($\tk,\tl$) are defined with respect to the reference vector $\vec{v}^\perp$,
whereas the third one $(\tkl)$ represents the relative angle between the two emitted partons in the 
transverse plane.
 
The matrix element of the RR contribution is proportional to the leading-order triple-collinear 
splitting functions~\cite{Campbell:1997hg,Catani:1998nv}, which contain overlapping divergences that 
cannot be disentangled with a single parametrisation. We then start from the following ansatz for the 
two-emission measurement function,
\begin{align}
{\cal M}_2(\tau; k,l) 
&= 
\exp 
\left[ 
-\tau q_T 
\left( 
\frac{q_T}{(1-x) P^-}
\right)^n
{\cal F}(a,b,z,t_k, t_l, t_{kl})
\right],
\end{align}
in which the observable is again described by the parameter $n$ and a function 
${\cal F}(a,b,z,t_k, t_l, t_{kl})$, whose explicit form can become quite lengthy. Even worse,
this function may vanish in the singular limits of the matrix element, and in this case one needs to 
apply sector-decomposition steps to correctly extract the associated divergences.

With the above form of the measurement function, one can easily follow the dependence on the variable 
$q_T$, which we integrate out analytically. In order to factorise the remaining phase-space divergences, we apply
a mixed strategy that consists of sector-decomposition steps, non-linear transformations and selector 
functions. This allows us to bring all divergences into monomial form, and we finally perform a Laurent 
expansion in the two regulators. For the numerical integration of the coefficients in this double expansion, 
we rely on {\tt pySecDec} \cite{Borowka:2017idc} and its {\tt Cuba} \mbox{implementation~\cite{Hahn:2004fe}}.

\section{Renormalisation}
\noindent
While the computation of the bare matching kernels can be performed in a universal framework for 
SCET-1 and SCET-2 observables, their renormalisation aspects are different, and we discuss them
one by one in this section.

\subsection{SCET-1 observables}
\noindent
In the combined Mellin-Laplace space the renormalisation of the beam-function matching kernels takes a 
multiplicative form, 
${\cal \widehat{I}}_{q\leftarrow j} = Z_q^B \sum_k{\cal \widehat{I}}_{q\leftarrow k}^{0} \,\widehat{Z}_{k\leftarrow j}^f$, 
where $Z_q^B$ subtracts the UV divergences of the beam function, whereas $\widehat{Z}_{k\leftarrow j}^f$ captures 
the IR divergences that match the UV divergences of the pdf. The renormalised matching 
kernels fulfil the renormalisation-group equation (RGE)
\begin{align}
\frac{d}{d\ln \mu} \; {\cal\widehat{I}}_{i\leftarrow j}(N,\tau,\mu)
=&
\left[ 2g(n) \,\Gamma_{\rm cusp}^{i}(\als) \, L +\gamma^{B}(\als)
\right] {\cal\widehat{I}}_{i\leftarrow j}(N,\tau,\mu)
\nonumber\\
&-2 \sum_{k} \,{\cal\widehat{I}}_{i\leftarrow k}(N,\tau,\mu) \,\widehat{P}_{k\leftarrow j}(N,\als) \,,
\label{eq:rge:scet1}
\end{align}
where $g(n) = (n+1)/n$,  $L=\ln \big( \mu\bar{\tau}/\left(q_-\bar{\tau}\right)^{1/g(n)}\big)$,  
$\Gamma_{\rm cusp}^{i}(\als)$ is the cusp anomalous dimensions in the representation of the parton $i$,  
$\gamma^{B}(\als)$  is the non-cusp anomalous dimension, and $\widehat{P}_{k\leftarrow j}(N,\als)$ are the 
DGLAP splitting functions in Mellin space. Expanding the anomalous dimensions in the form 
$  G(\alpha_s) = \sum_{m=0}^{\infty} G_m \,(\frac{\alpha_s}{4\pi})^{m+1}$, the two-loop solution of the RGE becomes,
\begin{align}
&{\cal\widehat{I}}_{i\leftarrow j}(N,\tau,\mu)  = 
\delta_{ij} + \left( \frac{\alpha_s}{4 \pi} \right) 
\bigg\{ \Big( g(n)\, \Gamma_0^i \,L^2 
+ \gamma_0^B \,L \Big) \delta_{ij}  
- 2L\,\widehat{P}_{i\leftarrow j}^{(0)}(N)
+ \widehat{I}_{i\leftarrow j}^{(1)}(N) \bigg\}
\\[0.2em]  
&\;
+\left( \frac{\alpha_s}{4 \pi} \right)^2 
\bigg\{ \bigg( \frac{g(n)^2\,(\Gamma_0^i)^2}{2} L^4 
+  g(n)\,\Gamma_0^i\Big( \gamma_0^B + \frac{2\beta_0}{3} \Big) 
L^3 + \Big( g(n)\,\Gamma_1^i + \frac12(\gamma_0^B)^2 
+\beta_0 \gamma_0^B  \Big) L^2  
+  \gamma_1^B L\bigg) \delta_{ij}
\nonumber\\[0.2em]  
&\qquad\qquad\quad
-2\Big( g(n)\,\Gamma_0^i L^3 + \big( \beta_0 + \gamma_0^B\big) L^2\Big) \widehat{P}_{i\leftarrow j}^{(0)}(N)
+ \Big( g(n)\,\Gamma_0^i  L^2 +  (\gamma_0^B +2\beta_0) L \Big)\,
\widehat{I}_{i\leftarrow j}^{(1)}(N)  
\nonumber\\[0.2em]  
&\qquad\qquad\quad
+2\sum_k \bigg( L^2\, \widehat{P}_{i\leftarrow k}^{(0)}(N) 
- L \, \widehat{I}_{i\leftarrow k}^{(1)}(N) \bigg) \widehat{P}_{k\leftarrow j}^{(0)}(N)  
- 2L\,\widehat{P}_{i\leftarrow j}^{(1)}(N)
+  \widehat{I}_{i\leftarrow j}^{(2)}(N) \bigg\} \,.
\nonumber
\end{align}
The renormalisation constants $Z_q^B$ and $\widehat{Z}_{k\leftarrow j}^f$ fulfil similar RGE that are controlled
by the first line or the second line of \eqref{eq:rge:scet1}, respectively. Their explicit form up to two loop-order 
can be found e.g.~in~\cite{Bell:2021dpb} for $Z_q^B$ and in \cite{Bell:2022nrj} for $\widehat{Z}_{k\leftarrow j}^f$.
From the pole terms of the bare matching kernels we then extract the non-cusp anomalous dimension $\gamma_m^B$, 
and from the finite terms we obtain the non-logarithmic coefficients $\widehat{I}_{i\leftarrow j}^{(m)}(N)$, 
which we sample for different values of the Mellin \mbox{parameter $N$}.

\subsection{SCET-2 observables}
\noindent
In the SCET-2 case we follow the collinear-anomaly approach \cite{Becher:2010tm,Becher:2011pf}, which states
that the product of the soft, collinear and anti-collinear functions can be refactorised in the form 
\begin{align}
&\left[
\widehat{\cal I}_{q\leftarrow i}(N_1,\tau,\mu,\nu) \;
\widehat{\cal I}_{\bar{q}\leftarrow j}(N_2,\tau,\mu,\nu) \; 
{\cal S}_{q\bar{q}}(\tau,\mu,\nu) 
\right]_{Q}
\nonumber\\
&\qquad=
\left( Q\taubar\right)^{-2F_{q\bar{q}}(\tau,\mu)} \;
\widehat{I}_{q\leftarrow i}(N_1,\tau,\mu) \;
\widehat{I}_{\bar{q}\leftarrow j}(N_2,\tau,\mu)\,,
\label{eq:refact}
\end{align}
where $Q^2=q_+ q_-$, and the quantities on the right-hand side are referred to as the 
collinear-anomaly exponent $F_{q\bar{q}}(\tau,\mu)$ and the refactorised matching kernels 
$\widehat{I}_{i\leftarrow j}(N,\tau,\mu)$. The former renormalises additively, 
$F_{q\bar{q}}^{0}=F_{q\bar{q}}+Z_{q\bar{q}}^F$, and it obyes the RGE
\begin{align}
\frac{d}{d\ln \mu} \; F_{q\bar{q}}(\tau,\mu) 
= 2\,\Gamma_{\rm cusp}^q(\als)\,,
\end{align}
which is solved by
\begin{align}
F_{q\bar{q}}(\tau,\mu)  &= 
\left( \frac{\alpha_s}{4 \pi} \right) 
\Big\{ 2\Gamma_0^q \,L 
+ d_1 \Big\}
+\left( \frac{\alpha_s}{4 \pi} \right)^2 
\Big\{ 2 \beta_0\Gamma_0^q\, L^2 
+ 2 \left( \Gamma_1^q + \beta_0 d_1 \right) L + d_2 \Big\},
\end{align}
where now  $L=\ln ( \mu\bar{\tau})$. The two-loop expression for the anomaly counterterm 
$Z_{q\bar{q}}^F$ can be found e.g.~in~\cite{Bell:2022nrj}.

The refactorised matching kernels, on the other hand, obey an RGE that is structually of the same form
as the one discussed in the previous section with $g(n)\to 1$ and $L\to\ln ( \mu\bar{\tau})$, while the
non-cusp anomalous dimension is usually expressed in terms of the collinear quark and gluon anomalous
dimensions in this case, i.e.~$\gamma^B(\als)\to-2\gamma^{i}(\als)$ with $i\in\{q,g\}$. As the latter are 
observable-independent, we use them to cross-check our calculation, similar to the cusp anomalous dimension 
$\Gamma_{\rm cusp}^{i}(\als)$. 
In essence we thus extract the non-logarithmic terms $d_m$ of the anomaly 
exponent and the coefficients $\widehat{I}_{i\leftarrow j}^{(m)}(N)$ of the refactorised matching 
kernels for SCET-2 observables.

\section{Results}
\noindent
With this setup, we computed the quark beam functions for transverse-momentum resummation and N-jettiness. 
As the NLO case is trivial, we focus here on the NNLO numbers, which we 
present in the form 
$\gamma_1^B = \gamma_1^{C_F} C_F^2 + \gamma_1^{C_A} C_F C_A + \gamma_1^{n_f} C_F T_F n_f$,
and similarly for $d_2$. For the non-logarithmic terms of the matching kernels we use the  
decomposition
\begin{align}
\widehat{I}_{q\leftarrow q}^{(2)}(N)&=
C_F^2 \; \widehat{I}_{q\leftarrow q}^{(2,C_F)}(N)
+ C_F C_A \; \widehat{I}_{q\leftarrow q}^{(2,C_A)}(N)   
+ C_F T_F n_f \; \widehat{I}_{q\leftarrow q}^{(2,n_f)}(N)
+ C_F T_F \; \widehat{I}_{q\leftarrow q}^{(2,T_F)}(N)\,,
\nonumber\\
\widehat{I}_{q\leftarrow g}^{(2)}(N) &=
C_F T_F \; \widehat{I}_{q\leftarrow g}^{(2,C_F)}(N) 
+ C_A T_F \; \widehat{I}_{q\leftarrow g}^{(2,C_A)}(N) \,,
\nonumber\\
\widehat{I}_{q\leftarrow \bar q}^{(2)}(N) &=
C_F (C_A - 2 C_F) \; \widehat{I}_{q\leftarrow \bar q}^{(2,C_{AF})}(N)  
+ C_F T_F \; \widehat{I}_{q\leftarrow q}^{(2,T_F)}(N) \,,
\nonumber\\
\widehat{I}_{q\leftarrow q'}^{(2)}(N) &=
\widehat{I}_{q\leftarrow \bar q'}^{(2)}(N) =
C_F T_F \; \widehat{I}_{q\leftarrow q}^{(2,T_F)}(N)\,.
\end{align}
There are thus seven independent coefficients to describe quark beam functions at NNLO, which we evaluate for
ten values of the Mellin parameter $N\in\{2,4,6,8,10,12,14,16,18,20\}$. 

\begin{table}[!t]
	\centering
	\renewcommand\arraystretch{1.4}
	\begin{tabular}{|c|c|c|}
		\hline
		$p_T$ &
		~~~~~Analytic~~~~~ &
		~~~~This work~~~~\\
		\hline
		\hline
		$d_2^{n_f}$ & 
		$ -8.296 $  & 
		$ -8.293(4)$  \\
		\hline
		$d_2^{C_A}$ &
		$ -3.732 $  & 
		$ -3.705(40)$  \\
		\hline
		$d_2^{C_F}$ &
		$ 0 $  & 
		$ 0.044(74)$  \\
		\hline
	\end{tabular}
	\hspace{10mm}
	\begin{tabular}{|c|c|c|}
		\hline
		N-jet&
		~~~~~Analytic~~~~~ &
		~~~~This work~~~~\\
		\hline
		\hline
		$\gamma_1^{n_f}$ &
		$-26.699$  & 
		$-26.700(34) $  \\
		\hline
		$\gamma_1^{C_A}$ &
		$-6.520$        & 
		$-6.549(324) $    \\
		\hline
		$\gamma_1^{C_F}$ &
		$21.220$    & 
		$21.223(518) $  \\
		\hline
	\end{tabular}
	\caption{\small{Two-loop anomaly coefficient for transverse-momentum resummation (left) and  two-loop 
			non-cusp anomalous dimension for N-jettiness (right). The analytic results have been extracted
			from~\cite{Gehrmann:2014yya,Gaunt:2014xga}.\\[-1.5em]}} 
	\label{tab:anomalousdimensions}
\end{table}

\begin{figure}[!h]
	\centerline{
		\includegraphics[width=0.45\textwidth]{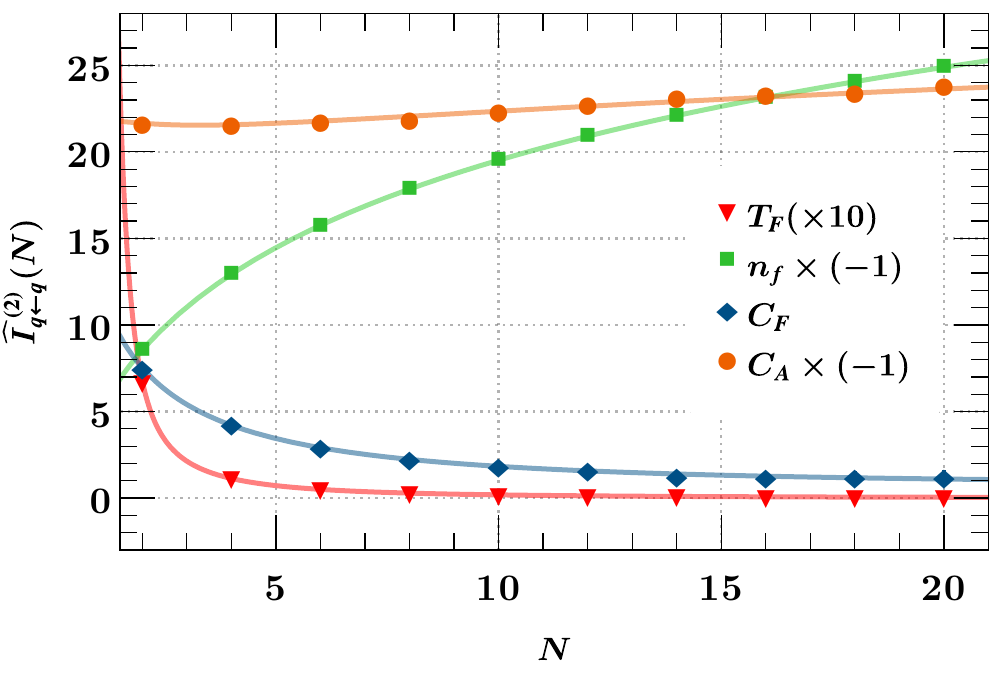}
		\includegraphics[width=0.473\textwidth]{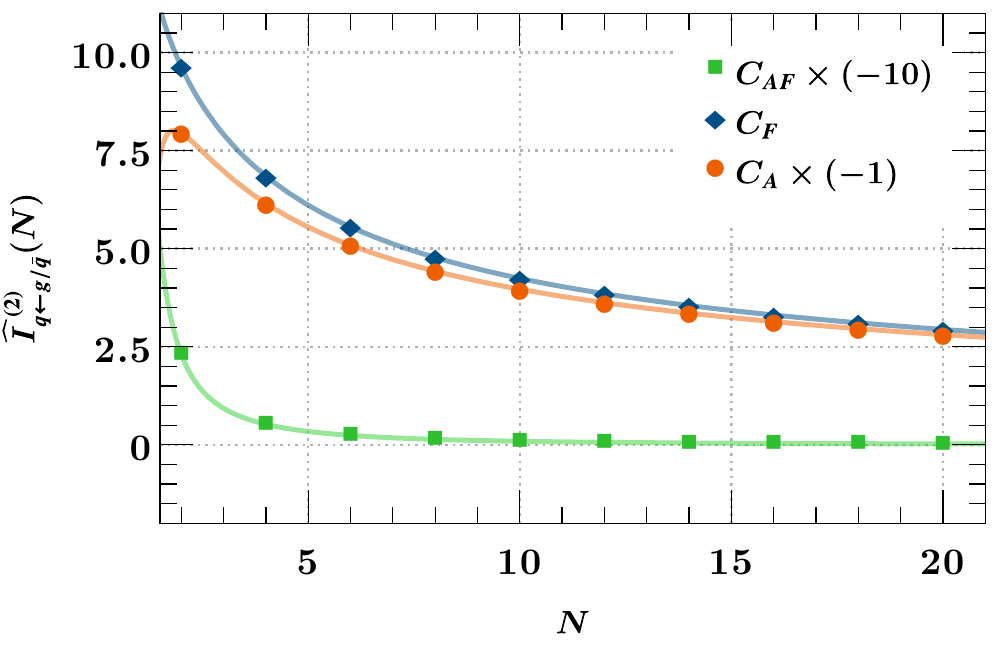}
			}
	\centerline{
	\includegraphics[width=0.45\textwidth]{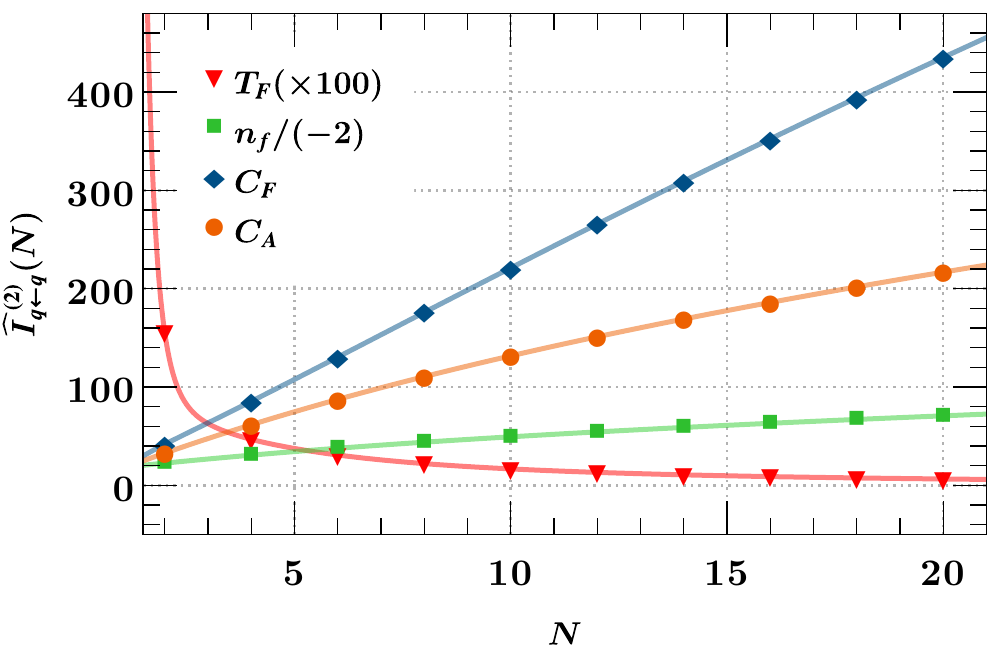}
	\includegraphics[width=0.473\textwidth]{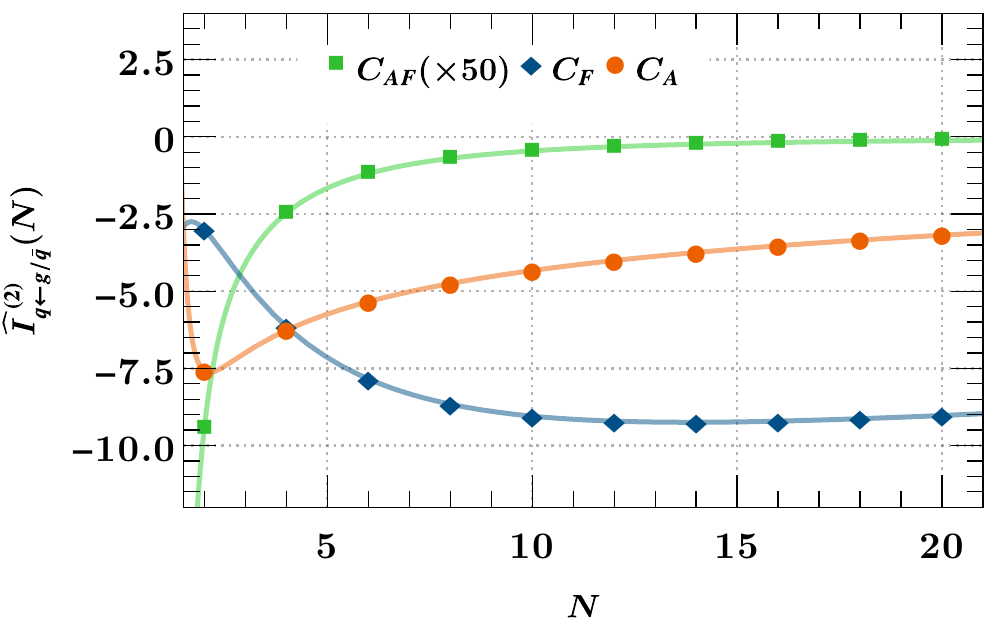}
}
	\caption{\small{
			Two-loop non-logarithmic matching kernels for transverse-momentum resummation (upper line)
			and N-jettiness (lower line). Left: The contributions to the diagonal kernels 
			$\widehat{I}_{q\leftarrow q}^{(2)}(N)$. Right: The same for the off-diagonal kernels 
			$\widehat{I}_{q\leftarrow g}^{(2)}(N)$ and $\widehat{I}_{q\leftarrow \bar q}^{(2)}(N)$. The solid 
			lines represent the analytical results from \cite{Gehrmann:2014yya,Gaunt:2014xga}, and the dots 
			show the numbers of our automated approach. The numerical uncertainties of the latter are too 
			small to be visible in the plots.}
	}
	\label{fig:pt-resummation}
\end{figure}

We start with transverse-momentum resummation, which is a SCET-2 observable and hence corresponds to the case $n=0$. According 
to the collinear-anomaly relation \eqref{eq:refact}, we need to combine the bare beam-function matching kernels 
with the corresponding soft function that we obtain from {\tt SoftSERVE}. Our results for the collinear-anomaly 
exponent $d_2$ are shown in Table~\ref{tab:anomalousdimensions}, and the non-logarithmic terms of the Mellin-space 
matching kernels are displayed in the upper panels of Figure~\ref{fig:pt-resummation}. The plots also show the 
analytic results from~\cite{Gehrmann:2014yya}, and we observe a very good agreement with the numerical predictions
of our novel automated approach.

We next consider the N-jettiness beam function, which is a SCET-1 observable with $n=1$. In this case we extract
the two-loop non-cusp anomalous dimension $\gamma_1^B$ and the non-logarithmic terms of the matching kernels 
$\widehat{I}_{i\leftarrow j}^{(2)}(N)$. The former are given in Table~\ref{tab:anomalousdimensions}, and the 
latter are displayed in the lower panels of Figure~\ref{fig:pt-resummation}. Our numbers are again in excellent
agreement with the known analytic results~\cite{Gaunt:2014xga}.

\section{Conclusion}
\noindent

We have presented a novel framework to calculate beam functions at two-loop order for a broad class of observables.
Our approach is based on suitable phase-space parametrisations and a systematic application of sector decomposition 
and non-linear transformations, which allows us to expose all phase-space singularities in an observable-independent
fashion. In order to avoid distributions, we furthermore chose to work in the Mellin-Laplace space, but we emphasise 
that our approach is not limited to this assumption, and we in fact already started to explore a direct calculation in  
momentum space. Our method has been implemented in the public package {\tt pySecDec}, and it has been validated against known results in the literature for transverse-momentum resummation and N-jettiness. As a further application, we recently calculated the quark beam function for jet-veto resummation~\cite{Bell:2022nrj}. In the future we plan to extend our
setup to gluon beam functions as well as to publish an automated standalone {\tt C++} code in the spirit of 
{\tt SoftSERVE}.

\subsection*{Acknowledgement}
\noindent
This work was supported by the Deutsche Forschungsgemeinschaft 
(DFG, German Research Foundation) under grant 396021762 - TRR 257 
(\emph{``Particle Physics Phenomenology after the Higgs Discovery''}).

\bibliographystyle{JHEP}
\bibliography{references}
\end{document}